\documentstyle[12pt]{article}
\begin{document}
\baselineskip=18pt
\begin{center}
{\large{\bf Decoherence of mesoscopic states of cavity fields}}
\end{center}
\vspace{0.5cm}
\baselineskip=12pt
\begin{center}
K. Fonseca Romero$^{\dag}$, M. C. Nemes$^{\dag}$, J. G. Peixoto
de Faria$^{\dag}$ 
\end{center}
\begin{center}
{\it Departamento de F\'{\i}sica, ICEX\\
     Universidade Federal de Minas Gerais\\
     C.P. 702,\ \ 30161-970 Belo Horizonte, M.G.,\ \ Brazil} 
\end{center}
\begin{center}
 and
\end{center}
\begin{center}
A. N. Salgueiro$^{\ddagger}$, A. F. R. de Toledo Piza
\end{center}
\begin{center}
{\it Instituto de F\'{\i}sica  \\
     Universidade de S\~ao Paulo \\
     CP 66318,\ \ 05315-970 S\~ao Paulo, S.P.,\ \ Brazil}
\end{center}
\baselineskip=15pt
\vspace{0.5cm}
\begin{center}
{\bf Abstract}
\end{center}
\vspace{0.5cm}

We show that two-atom correlation measurements of the type involved in
a recent experimental study of the evolution of a mesoscopic
superposition state prepared in a definite mode of a high-Q cavity can
be used to determine the eigenvalues of the reduced density matrix of
the field, provided the assumed dynamical conditions are actually
fulfilled to experimental accuracy. These conditions involve i) a
purely dispersive coupling of the field to the Rydberg atoms used to
manipulate and to monitor the cavity field, and ii) the effective
absence of correlations in the ground state of the system consisting
of the cavity coupled to the ``reservoir'' which accounts for the
decoherence and damping processes. A microscopic calculation at zero
temperature is performed and compared to master equation results.

\vspace{0.5cm}
\begin{center}
{\it As of October 20, 1997.}

\today
\end{center}

\vspace{1cm}
\noindent PACS numbers:

\vspace{\fill}

\noindent\makebox[66mm]{\hrulefill}

\footnotesize 

$^{\dag}$Supported partly by Conselho Nacional de Desenvolvimento
Cient\'{\i}fico e Tecnol\'ogico (CNPq), Brazil.

$^{\ddagger}$Supported by Funda\c{c}\~ao de Amparo \`a Pesquisa do
Estado de S\~ao Paulo (FAPESP), Brazil.

\normalsize
\newpage
\baselineskip=15pt

The conception of an experimental setup to produce a superposition of
mesoscopic quantum states and to observe its subsequent decoherence
\cite{omn} in microwave cavities has been preceded by a series of
meticulous experimental and theoretical studies
\cite{roch95}-\cite{prop}.  Several schemes have been proposed, based
on two atom correlation measurements involving circular Rydberg atoms,
and the actual implementation of one such scheme has been recently
reported \cite{karou}. It can be broadly pictured as involving two
successive stages, which consist respectively in the preparation of
the desired superposition state and in the subsequent monitoring of
its time evolution under influence of the unavoidable loss processes
present in the cavity. As the last step in the preparation stage, what
is now often referred to as the ``reduction postulate'' \cite{voneu}
is used instrumentally in an essential way to select the desired state
of the field (see below). The subsequent stage, however, involves just
strictly causal quantum evolution laws, and is therefore, in
principle, amenable to a complete theoretical description, including
the coherence loss of the prepared superposition state \cite{roch95,
prop}. In particular, as it is appropriate for bona-fide physical
properties and processes, the coherence loss should be characterized
and described in a basis independent fashion. We show in continuation
that this can be done in a particularly simple way in the present
case, given the specific properties of the prepared superposition
state, and no more than ordinary assumptions concerning the external
couplings underlying the mechanism of coherence loss. More
specifically, we show that, under the experimental conditions stated
in \cite{karou}, two-atom correlations of the type involved in the
measurements reported there can be used in order to determine
completely the eigenvalues of reduced density matrices which describe
the possible quantum states of the cavity field following the
preparation stage. The time dependence of the observed two-atom
correlations subsequently to the preparation of the superposition
state constitutes therefore a measurement of intrinsic features of the
evolution of the field which bear directly on the coherence loss
process. Moreover, to the extent that this evolution depends both on
the nature of the initial state and on dynamical assumptions
concerning the dissipative process, through an analysis of the type
proposed here such measurements can provide {\it experimental} checks
on the actual outcome of the preparation stage and on the adequacy of
the assumed dissipative dynamics, at no higher cost than a less
inclusive analysis of the two-atom correlation data.

The experiment reported in \cite{karou} involves a high-$Q$ cavity $C$
located between two low-$Q$ cavities (Ramsey zones $R_1$ and $R_2$)
fed with classical fields which steer the internal state of the
crossing circular Rydberg atoms. The transition between two near
atomic levels, usually denoted as $\mid e \rangle$ and $\mid g
\rangle$, is resonant with the fields in cavities $R_1$ and
$R_2$. Each atom is initially prepared in the state $\mid e
\rangle$. After leaving $R_1$ it is in a superposition of the states
$\mid e \rangle$ and $\mid g \rangle$. The fields in $R_1$ and in
$R_2$ (and their relative phases) are chosen so that their action on
the atoms is given in both cases by the unitary transformation

\[
U_{R}:\;\;\; \mid e \rangle\rightarrow\frac{1}{\sqrt{2}}\left(\mid e
\rangle + \mid g\rangle\right);\;\;\; \mid g \rangle \rightarrow
\frac{1}{\sqrt{2}}\left(-\mid e \rangle + \mid g\rangle\right).
\]

\noindent The high-$Q$ cavity $C$ stores a coherent field $\mid\alpha
\rangle$. The coupling between the atom and the field in this cavity
is measured by the Rabi frequency $\Omega$. Two situations will be
considered: a) the frequency $\omega$ of a relevant mode of $C$ is
nearly resonant with the transition frequency $\omega_{ie}$ connecting
$\mid e\rangle$ to a third state $\mid i\rangle$ (assumed for
definiteness to lie above $\mid e\rangle$ in energy) and is far off
resonance with any transitions involving level $\mid g\rangle$
\cite{prop}; and b) $\omega$ is nearly resonant with the transition
frequency $\omega_{eg}$ connecting $\mid e\rangle$ to $\mid g\rangle$
\cite{roch95,karou}. In both cases the detuning $\delta$ is assumed
large enough so that real transitions are effectively hindered and the
atom-field interaction leads essentially to $1/\delta$ dispersive
frequency shifts. In this way, the atom-field coupling during a time
$t$ produces an atomic level dependent dephasing of the field and thus
generates entangled states of the atom and field. This can be
represented by the unitary operators

\begin{eqnarray}
\label{pertmod}
{\rm case}&&{\rm a:}\;\;\;\;\;\;U_C^{(a)}=e^{-i\phi a^\dagger a}\mid
e\rangle \langle e\mid + \mid g\rangle\langle g\mid \nonumber \\
\\{\rm case}&&{\rm b:}\;\;\;\;\;\;U_C^{(b)}=
e^{i\phi(a^\dagger a+1)} \mid e\rangle \langle e\mid + e^{-i\phi
a^\dagger a} \mid g\rangle\langle g\mid \nonumber
\end{eqnarray}

\noindent with $\phi(t)=\Omega^2 t/\delta$. In these expressions the
operators $a^\dagger$, $a$ create and annihilate photons in the
relevant mode of $C$. The effect of these operators on a coherent
state of this cavity amounts thus simply to a change of {\it phase}
which depends on the state of the intervening two-level atom, so that
the resulting state of the field can be expressed in terms of just two
coherent states. It should be emphasized that this feature depends
crucially on the validity of the purely dispersive limit mentioned
above, which requires values of $\Omega/\delta$ sufficiently small to
ensure the validity of the implied low-order perturbative treatment
\cite{priv}.

This idealization of the involved dynamics allows one to describe the
experiment in a symbolically compact form. When an atom is sent
through the apparatus, the change in the state of the combined system
consisting of atom plus field is described by $U_R U_C U_R$, so that
for a factorized initial state $\rho^F \otimes\mid e\rangle\langle
e\mid$ (with $\rho^F=\mid\alpha \rangle\langle\alpha\mid$) one has

\[
U_R U_C U_R(\rho^F \otimes\mid e\rangle\langle e\mid)U_R^\dagger
U_C^\dagger U_R^\dagger
\]

\noindent for the combined system. The field state resulting from the
final observation of the atom in state $\mid e\rangle$ or $\mid
g\rangle$ is obtained by replacing the product of unitary operators
respectively by the reduced operators

\begin{equation}
\label{scrpu}
{\cal U}_{^e_g}\equiv\langle ^e_g\mid U_R U_C U_R \mid e\rangle
\end{equation}

\noindent (or their adjoints) with an appropriate trace normalization
factor. This is what we referred to as the preparation stage of the
experiment. The instrumental role attributed to the ``reduction
postulate'' in this stage appears explicitly in this last step. One
gets, for the two cases considered,

\begin{equation}
\label{instates}
{\cal U}_{^e_g}^{(a)}=\frac{1}{2}\left[e^{-i\phi a^\dagger a}\mp
1\right]\;\;\;\;\;\;{\rm and}\;\;\;\;\;\; 
{\cal U}_{^e_g}^{(b)}=\frac{1}{2}\left[e^{i\phi(a^\dagger a +1)}\mp
e^{-i\phi a^\dagger a}\right].
\end{equation}

\noindent The indices $e,g$ and the double signs refer respectively to
the final observation of the atom in states $\mid e\rangle$ and $\mid
g\rangle$.

We now turn to the second stage of the two atom correlation
measurement. The second atom is sent through the system following the
detection of the first. The passage of this second atom trough
the apparatus and its detection can be described using the same formal
tools adopted for the first atom. Let $\rho_{^e_g}^F$ be the trace
normalized reduced densities which describe the state of the field in
$C$ after the detection of the first atom in $e$ or $g$. The
conditional probabilities for the various possible results obtained in
the final measurement of the state of the second atom, for given
outcome of the measurement of the final state of the first atom, are
respectively given by

\begin{eqnarray}
\label{condprobs}
P_{ee}=Tr\left[{\cal{U}}_e^\dagger{\cal{U}}_e\rho_e^F\right]; &&
P_{ge}=Tr\left[{\cal{U}}_e^\dagger{\cal{U}}_e\rho_g^F\right] \nonumber
\\ \\
P_{eg}=Tr\left[{\cal{U}}_g^\dagger{\cal{U}}_g\rho_e^F\right]; &&
P_{gg}=Tr\left[{\cal{U}}_g^\dagger{\cal{U}}_g\rho_g^F\right]. \nonumber
\end{eqnarray}

\noindent Note that in e.g. $P_{ge}$ the first atom is detected in
state $g$ and the second in state $e$. Because of the assumed
normalization one has $P_{ee}+P_{eg}= 1=P_{ge}+P_{gg}$, and the
correlation signal measured in ref. \cite{karou} can be written simply
as

\begin{equation}
\label{eta}
\eta=P_{ee}-P_{ge}=Tr\left[{\cal{U}}_e^\dagger {\cal{U}}_e \left(
\rho_e^F-\rho_g^F\right)\right].
\end{equation}

\noindent The reduced densities appearing here refer to the
state of the field in $C$ at the time of the passage of the second
atom through the apparatus. On the one hand, these expressions connect
the two-atom correlation data directly with theoretical models for the
fully causal time evolution of $\rho_{^e_g}^F$ during the time lapse
between the passage of the two atoms. This involves in fact the only
truly ``complex'' aspect of the dynamics, which is responsible for the
dissipative evolution of the field state in $C$. Since moreover one
can always express the reduced densities in terms of their (time
dependent) eigenvectors and eigenvalues, Eqs. (\ref{condprobs}) and
(\ref{eta}) also show directly that measured values of the various
conditional probabilities and of $\eta$ can be expressed in terms of
these same ingredients. As shown below, the eigenvalues are in fact
{\it determined} by $P_{ee}$ and $P_{ge}$, {\it provided} specific but
ordinary assumptions concerning the dissipation mechanism are made.
In this sense, they provide a basis independent characterization of
the decoherence process much as quantities based on more general
constructs which are nonlinear in $\rho$ such as e.g. the
``idempotency defect'' $1-Tr\rho^2$ \cite{pert} or entropy-like
functionals of the density \cite{voneu}. The assumptions concerning
the damping mechanism will allow, in particular, for an {\it exact}
treatment, independently of the also customary recourse to master
equation approximations.

In order to model the dissipative effects in $C$ during the time
between the passage of the two atoms, we introduce as usual a set
of harmonic oscillators with frequencies $\omega_k$ and
creation/annihilation operators written as $b_k^\dagger$, $b_k$ (the
``heat bath'') coupled to the relevant mode of the cavity through the
bilinear expression

\begin{equation}
\label{rwa}
H_{int}=\sum_k\hbar\gamma_k (a^\dagger b_k + b_k^\dagger a).
\end{equation}

\noindent This differs from the also often used position-position
coupling by the omission of ``anti-resonant'' terms involving
products of two creation and of two annihilation operators. These terms
are actually expected to contribute very little to the damping process
in view of the smallness of the ratio $\gamma_k/\omega$ in the present
context. The resulting system of coupled oscillators has the well
known properties that i) its ground state is the product of the ground
states of the different oscillators (no ground state correlations) and
ii) an initial state written as a product of coherent states, i.e.

\begin{equation}
\label{prdstate}
\mid t=0\rangle=\mid\alpha\rangle\prod_k\mid\beta_k\rangle
\end{equation}

\noindent retains this form at all times, the amplitudes $\alpha(t)$ and
$\beta_k(t)$ being solutions to the linear coupled equations

\begin{eqnarray}
\label{coupleqs}
i\dot{\alpha}&=&\sum_k\gamma_k\beta_k \nonumber \\
\\
i\dot{\beta_k}&=&(\omega_k-\omega)\beta_k+\gamma_k\alpha \nonumber
\end{eqnarray}

\noindent with the given initial conditions, in an interaction picture
where the time dependence associated with the constant of motion
$\hbar\omega\left(a^\dagger a+\sum_k b^\dagger_k b_k\right)$ has been
eliminated. This solution is in fact the relevant one when we then
take as initial states the states generated by the action of the
operators obtained in Eq. (\ref{instates}) on a coherent state
$\mid\alpha\rangle$ and a heat bath at temperature zero. Since the
state of the relevant mode of $C$ is then the sum of two linearly
independent (though, strictly speaking, never orthogonal) coherent
states, we see that we will have at all times a sum of two terms, each
being of the type shown in Eq. (\ref{prdstate}). Consequently, the
reduced density describing the state of the field in $C$ will have its
trace exhausted in a (time dependent) two-dimensional subspace of the
corresponding oscillator space \cite{ric}. This particular feature is
in fact retained in the currently used master equation description of
the damping process (see e.g. \cite{prop}), which however gives
quantitatively different results both at very short times (where it
gives a linear rather than a quadratic time dependence for decoherence
measures such as the idempotency defect \cite{pert}) and for long
times in the asymptotically damped region.

The computationally simplest case to consider corresponds to case a) as
defined above, with $\phi=\pi$. The ``initial'' state (after the
detection of the first atom) is then

\[
\mid t=0\rangle=N_\alpha\left[\mid\alpha\rangle \mp
\mid -\alpha\rangle\right]\otimes\mid 0_b\rangle
\]

\noindent where $\mid 0_b\rangle$ stands for the ``bath'' ground state
(zero temperature), the double sign corresponds to the detection of
the first atom in $e$ or $g$ and $N_\alpha$ is the appropriate
normalization factor. The solution of Eqs. (\ref{coupleqs}) then gives
at time $t$

\[
\mid t\rangle=N_\alpha\left[\mid\alpha(t)\rangle\prod_k\mid
\beta_k(t)\rangle \mp \mid -\alpha(t)\rangle\prod_k\mid
-\beta_k(t)\rangle\right].
\]

\noindent It is then a simple matter to obtain the $C$-field reduced
density at time $t$ as

\begin{equation}
\label{rho}
\rho_{^e_g}^F(t)=N_\alpha^2\left[\mid\alpha(t)\rangle\langle\alpha(t)
\mid + \mid -\alpha(t)\rangle\langle -\alpha(t)\mid\mp\Gamma_b(t)
\left(\mid\alpha(t)\rangle\langle -\alpha(t)\mid + \mid
-\alpha(t)\rangle\langle \alpha(t)\mid\right)\right]
\end{equation}

\noindent where the real function $\Gamma_b(t)$, which contains the
decoherence effects of the bath oscillators, is given by

\[
\Gamma_b(t)=\prod_k\langle -\beta_k(t)\mid\beta_k(t)\rangle.
\] 

\noindent The non-vanishing eigenvalues $\lambda_\pm(t)$ of each of
these two matrices are 

\begin{equation}
\label{eigena}
\lambda^{(^e_g)}_+(t)=\frac{\left(1+\Gamma_a(t)\right)\left(1\mp
\Gamma_b(t)\right)}{2\left(1\pm\Gamma_a(0)\right)};\;\;\;\;
\lambda^{(^e_g)}_-(t)=\frac{\left(1-\Gamma_a(t)\right)\left(1\pm
\Gamma_b(t)\right)}{2\left(1\pm\Gamma_a(0)\right)}
\end{equation}

\noindent where $\Gamma_a(t)=\langle-\alpha(t)\mid\alpha(t)\rangle$,
with associated eigenvectors

\[
\mid\lambda_\pm(t)\rangle\;\;\propto\;\;\mid\alpha(t)\rangle\pm\mid
-\alpha(t)\rangle.
\] 

\noindent Since in this special case the operator $\exp(i\pi a^\dagger
a)$ appearing in ${\cal{U}}_e^\dagger{\cal{U}}_e$ is simultaneously
diagonal (with eigenvalues $\pm1$ for $\mid\lambda_\pm(t)\rangle$)
with the reduced densities, the correlation signal $\eta(t)$ measured
in \cite{karou} assumes the simple form

\[
\eta(t)=\lambda^{(e)}_-(t)-\lambda^{(g)}_-(t).
\]

\noindent To the extent that the overlap $\Gamma_a(0)$ is and
$\Gamma_a(t)$ remains negligible, it is clear from Eqs. (\ref{eigena})
that $\lambda^{(^e_g)}_-\approx\lambda^{(^g_e)}_+$, so that the above
result reduces to

\begin{equation}
\label{smov}
\eta(t)\approx\lambda^{(g)}_+(t)-\lambda^{(g)}_-(t)=1-2
\lambda^{(g)}_-(t) \approx
\lambda^{(e)}_-(t)-\lambda^{(e)}_+(t)=1-2\lambda^{(e)}_+(t)
\end{equation}

\noindent which, together with the trace normalization, is sufficient
to determine the eigenvalues in this small overlap limit. One sees
moreover that, independently of the small overlap condition, the
eigenvalues of the reduced densities are determined by the separately
measured quantities $P_{ee}$ and $P_{ge}$ as $\lambda_-^{(e)}=P_{ee}$
and $\lambda_-^{(g)}=P_{ge}$.

Exactly the same analysis applies in this case when a master equation
approximation is used to describe the damping effects associated with
Eqs. (\ref{coupleqs}), as done e.g. in ref. \cite{prop}. In fact, the
master equation can be solved in closed form giving for the reduced
density an expression identical in form to Eq. (\ref{rho}) with

\[
\alpha(t)\rightarrow\alpha(0)e^{-\gamma t/2}\;\;\;\;\;\;{\rm{and}}
\;\;\;\;\;\; \Gamma_b(t)\rightarrow e^{-2|\alpha(0)|^2(1-e^{-\gamma
t})}
\]

\noindent where $\gamma=1/t_c$, the inverse of the damping time of the
field intensity in $C$.

Case b), which corresponds to the conditions of the experiment
reported in \cite{karou}, can be handled in precisely the same way as
case a), since the state of the field prepared in $C$ (i.e.,
immediately after the detection of the first atom) is still of the
form of a superposition of {\it just two} coherent states, namely

\[
\mid t=0\rangle=N_{\alpha\phi}\left[e^{i\phi}\mid\alpha
e^{i\phi}\rangle+\mid\alpha e^{-i\phi}\rangle\right]
\otimes\mid 0_b\rangle.
\]

\noindent Expressions for the two non-vanishing eigenvalues and for the
corresponding eigenvectors of the reduced densities can still be
calculated in a straightforward way, in terms of the appropriate
solutions of Eqs. (\ref{coupleqs}), taking into account the
non-orthogonality of the coherent state representation through the
appropriate overlap matrix. Since they are considerably more
cumbersome than in the previous case and will not be needed explicitly
in the following development we do not give them in full
here. 

Before turning to a discussion of the general case, we consider the
simpler expressions appearing in the small overlap limit
(i.e. $\mid\langle\alpha(t)e^{i\phi}
\mid\alpha(t)e^{-i\phi}\rangle\mid\ll 1$ so that the overlap matrix
essentially reduces to the unit matrix). Although the operators
involved in ${\cal{U}}_e^\dagger {\cal{U}}_e$ are no longer
simultaneously diagonal with the reduced densities, one still gets for
the correlation signal a result similar to Eq. (\ref{smov}), which can
be written as

\[
\eta(t)\approx\cos\left(\sum_k|\beta_k(t)|^2\sin 2\phi\right)
(\lambda_+-\lambda_-)
\]

\noindent where $\lambda_\pm\approx\lambda_\pm^{(^g)}\approx
\lambda_\mp^{(e)}$ stand for the small overlap approximations to the
eigenvalues of the reduced densities 

\[
\lambda_\pm=\frac{1\pm\mid\Gamma_b(t,\phi)\mid}{2}
\]

\noindent  with

\[
\Gamma_b(t,\phi)=e^{-2\sum_k|\beta_k(t)|^2 \sin^2\phi}
e^{i\theta(t)}\;\;\;\;\;\;{\rm and}\;\;\;\;\;\;
\theta(t)=\sum_k|\beta_k(t)|^2 \sin 2\phi.
\]

\noindent In the last equations $\alpha(t)$ and $\beta_k(t)$ stand
for the solutions to Eqs. (\ref{coupleqs}) with initial conditions
$\alpha(0)$, $\beta_k(0)=0$. This makes all the dependence on $\phi$
explicit. The possibility to express the small overlap approximation
to $\eta(t)$ in terms of reduced density eigenvalues is due to the
fact that in this limit the two densities $\rho^F_e$ and $\rho^F_g$
essentially commute and also have essentially the same eigenvalues,
i.e.

\[
\rho^F_e\approx\mid\lambda_+\rangle\lambda_+\langle\lambda_+\mid +
\mid\lambda_-\rangle\lambda_-\langle\lambda_-\mid\;;\;\;\;\;\;\;\;
\rho^F_g\approx\mid\lambda_+\rangle\lambda_-\langle\lambda_+\mid +
\mid\lambda_-\rangle\lambda_+\langle\lambda_-\mid
\]

\noindent with approximate eigenstates

\[
\mid\lambda_\pm\rangle\propto\pm e^{i(\theta(t)+\phi)}\mid\alpha(t)
e^{i\phi}\rangle+\mid\alpha(t)e^{-i\phi}\rangle.
\]

The correlation signal measured in \cite{karou}, Eq. (\ref{eta}), thus
essentially determines the eigenvalues of the reduced densities
$\rho^F_e$ and $\rho^F_g$ {\it in the small overlap limit}. Essential
for this result is the pairwise equality of the eigenvalues of the two
matrices in this limit. We then finally consider the general case and
show that, independently of the small overlap assumption, separate
measurements of $P_{ee}$ and $P_{ge}$ again determine the eigenvalues
of the trace normalized matrices $\rho^F_e$ and $\rho^F_g$. In fact,
it is easy to show that these quantities provide in general
independent linear combinations of the two non-vanishing eigenvalues
in each case, with coefficients which depend on the dynamics of the
atom-field interaction as expressed by matrix elements of the reduced
operators ${\cal{U}}_{^e_g}$, Eq.(\ref{scrpu}), in the reduced density
eigenvectors. In order to see this explicitly, it suffices to write
the reduced densities in diagonal form as

\begin{equation}
\label{diag}
\rho^F_{^e_g}=\mid\lambda^{(^e_g)}_+(t)\rangle \lambda^{(^e_g)}_+(t)
\langle\lambda^{(^e_g)}_+(t)\mid + \mid\lambda^{(^e_g)}_-(t)\rangle
\lambda^{(^e_g)}_-(t) \langle\lambda^{(^e_g)}_-(t)\mid
\end{equation}

\noindent which gives for the correlation functions

\[
P_{^{ee}_{ge}}=\langle\lambda^{(^e_g)}_+(t)\mid{\cal{U}}_e^\dagger
{\cal{U}}_e \mid\lambda^{(^e_g)}_+(t)\rangle \lambda^{(^e_g)}_+(t) +
\langle\lambda^{(^e_g)}_-(t)\mid{\cal{U}}_e^\dagger 
{\cal{U}}_e \mid\lambda^{(^e_g)}_-(t)\rangle \lambda^{(^e_g)}_-(t).
\]

\noindent These general expressions hold of course also in case a)
above, where one has

\[
\langle\lambda^{(^e_g)}_+\mid {\cal{U}}_e^\dagger {\cal{U}}_e
\mid\lambda^{(^e_g)}_+\rangle=0\;\;\;\;\;\;{\rm and}\;\;\;\;\;\;
\langle\lambda^{(^e_g)}_-\mid {\cal{U}}_e^\dagger {\cal{U}}_e
\mid\lambda^{(^e_g)}_-\rangle=1.
\]

\noindent As a result of the fact that in case a) the reduced
densities are simultaneously diagonal with the reduced operators
${\cal{U}}_{^e_g}$, the determination of the eigenvalues becomes in
fact independent of the structure of the (in this case, common)
eigenvectors.

We conclude from these results that the validity of the purely
dispersive coupling assumption of the two-level atom to the field in
cavity $C$, which leads to the validity of Eqs. (\ref{pertmod}), {\it
and} the adequacy of a coupling of the form of Eq. (\ref{rwa}) to
implement the dissipative dynamics allows for a direct experimental
monitoring of the intrinsic decoherence processes undergone by the
field as a result of its coupling with the ``heat bath''. This can be
done simply through the separate measurement of the conditional
probabilities $P_{ee}$ and $P_{ge}$, since they fully determine the
non-vanishing {\it eigenvalues} (in this case only two, for all times
after the preparation of the superposition state) of each one of the
reduced densities $\rho^F_e$ and $\rho^F_g$. Under less stringent
conditions, allowing for measurable effects of higher order
corrections (in $\Omega/\delta$) to Eqs. (\ref{pertmod}), one will
generate field states, after the passage of the first atom, which
cannot be reduced to superpositions of just two linearly independent
states (see e.g. ref. \cite{cats}), and will consequently have reduced
densities having a correspondingly higher number of non-vanishing
eigenvalues. The representation corresponding to Eq. (\ref{diag}) will
then have more than just two terms so that the {\it complete}
determination of intrinsic decoherence properties of the field will
require fancier measurements.

\end{document}